\documentclass[11pt]{article}
\usepackage{pennames}
\usepackage{moriond,epsfig}

\def\Journal#1#2#3#4{{#1} {\bf #2}, #3 (#4)}

\def\EPJ{{\em Eur. Phys. J.} C}

\def\NIMA{{\em Nucl. Instrum. Methods} A}

\def\PLB{{\em Phys. Lett.}  B}

\def\PBs{\relax\ifmmode{{\rm B}^0_{\rm s}}%
           \else${\rm B}^0_{\rm s}$\fi}%
\def\PaBs{\relax\ifmmode{\overline{\rm B}^0_{\rm s}}%
           \else${\rm B}^0_{\rm s}$\fi}%
\def\PsD{\relax\ifmmode{{\rm D}_{\rm s}^\pm}%
           \else${\rm D}_{\rm s}^\pm$\fi}%
\def\PsDst{\relax\ifmmode{{\rm D}_{\rm s}^{(\ast)\pm}}%
           \else${\rm D}_{\rm s}^{(\ast)\pm}$\fi}%
\def\PsDstm{\relax\ifmmode{{\rm D}_{\rm s}^{(\ast)-}}%
           \else${\rm D}_{\rm s}^{(\ast)-}$\fi}%
\def\PsDstp{\relax\ifmmode{{\rm D}_{\rm s}^{(\ast)+}}%
           \else${\rm D}_{\rm s}^{(\ast)+}$\fi}%

\def\mevcc{MeV/$c^2$}
\def\dmd{\ensuremath{\Delta m_\mathrm{d}}}
\def\dms{\ensuremath{\Delta m_\mathrm{s}}}

\def\be{\begin{equation}}
\def\ee{\end{equation}}
\def\bea{\begin{eqnarray}}
\def\eea{\end{eqnarray}}

\begin{document}
\vspace*{4cm}
%\title{NEW RESULTS IN THE SEARCH FOR $\PBs-\PBs$ OSCILLATIONS}
\title{HEAVY FLAVOURS AT LEP}

\author{ A. SCIAB\`A }

\address{INFN -- CNAF, via Berti-Pichat 6/2,\\
40127 Bologna, Italy}

\maketitle\abstracts{
This paper describes recent developments in Heavy Flavour physics at LEP,
focusing on a new result from the ALEPH Collaboration on
the search for flavour oscillations of the $\PBs$ meson. The impact of this
analysis on the world combination and the resulting limit on the $\PBs$
oscillation frequency are discussed.}

\section{Introduction}
Six years after the end of the LEP1 data taking, the LEP collaborations
continue producing impressive results in Heavy Flavours physics. Now very
near to its completion, the LEP Heavy Flavours research programme achieved
a very good precision in measuring some of the experimental quantities that
enter in the determination of the CKM matrix elements, like the b hadron
lifetimes, inclusive and exclusive b semileptonic branching ratios, or the
$\PBz$ oscillation frequency, reaching in some cases accuracies at the
percent level and significantly better than the theoretical uncertainties
involved in the CKM elements extraction~\cite{bphysics}.
Rather than reviewing these
results, the goal of this paper is to focus on a very recent result from the
ALEPH collaboration on the search for $\PBs$ oscillations, which represents
the most relevant development in Heavy Flavour physics from LEP during
the last few months.

The $\PBs$ flavour mixing is expected to happen exactly as in the case of
$\PKz$ and $\PBz$ mesons, as a  consequence of the flavour non-conservation
in charged weak-current interactions. However, a measurement of the $\PBs$
oscillation frequency $\dms$ is still lacking, because it is too high to be
resolved with the present experimental resolutions and statistics.

The interest in this measurement lies in the possibility to constrain much
more tightly the value of $V_{\mathrm{td}}$; in fact, using the
$\PBz$ oscillation frequency $\dmd$ alone, the accuracy in the
$V_{\mathrm{td}}$ determination is about 15\%, due the
uncertainty on the non-perturbative QCD calculations involved, while in
the $\dmd/\dms$ ratio this uncertainty largely
cancels~\cite{theor}.

If the width difference between the two $\PBs$ mass eigenstates and CP
violation effects are neglected, the probability for a $\PBs$ meson
produced at $t=0$ in a definite flavour eigenstate to decay at a time $t$
in the opposite (or same) flavour eigenstate can be simply expressed as:
\begin{equation}
  P(t)_{\PBs\rightarrow\PaBs(\PBs)}=\frac{1}{2\tau_s}
  e^{-\frac{t}{\tau_s}}\left[1\mp A\cos(\dms t)\right],
  \label{eq:prob}
\end{equation}
where $\tau_s$ is the average $\PBs$ lifetime, and $A\equiv1$.

In principle, using Eq.~\ref{eq:prob}, it is possible to measure
$\dms$ provided that,
for every selected $\PBs$ meson, one can measure $a)$ the flavour state
at production time, $b)$ the flavour state at decay time, and $c)$ the
proper decay time. If the experimental sensitivity is not enough to
perform such a measurement, the so-called
``amplitude method'', consisting in fitting the
amplitude $A$ of the oscillating term in Eq.~\ref{eq:prob} for a fixed value
of the frequency,
is commonly used to show the agreement of the data with
a certain hypothesis for the value of $\dms$~\cite{amplitude,duccio}; to do
this, the fitted value of $A$ is plotted as a function of the
oscillation frequency being tested, $\omega$.
It is expected for $A$ to be consistent with one at the true value of
$\dms$ and with zero at far lower frequencies.

The lower limit on $\dms$ at 95\% C.L. is
defined as the smallest value of $\dms$ for which an amplitude $A=1$
is not excluded at 95\% C.L., while the sensitivity of a measurement is
conventionally defined as the expected 95\% C.L. lower
limit on $\dms$ if the true value of $\dms$ were infinite.

\section{New experimental results}
In this section, the most recent experimental results on the
search for $\PBs$ oscillations from the ALEPH collaboration will
be described~\cite{newbs}.

Three different event selections have been applied to approximately four
million hadronic Z decays,
collected by the ALEPH detector~\cite{aleph,perf} from 1991 to 1995, and,
for the first analysis alone, to about 400000 hadronic Z decays taken at the
Z peak for calibration purposes from 1996 to 2000. The first selection
reconstructs exclusively some $\PBs$ hadronic decays; the second selection
reconstructs semileptonic $\PBs$ decays in a semi-exclusive manner, while
the third is an inclusive selection of semileptonic $\PBs$ decays.

\begin{figure}
  \begin{center}
  \epsfig{file=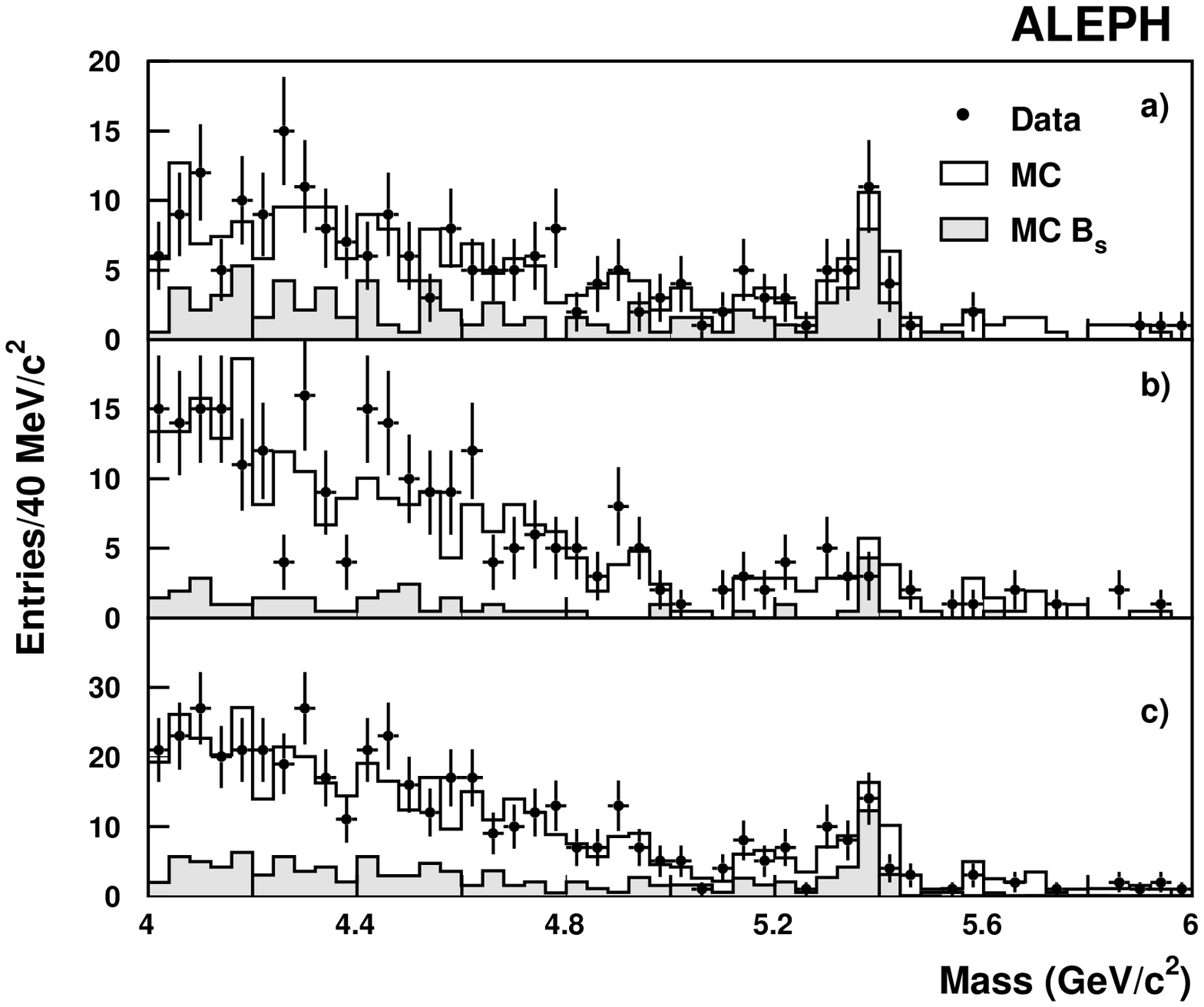,height=8.9cm}
  \epsfig{file=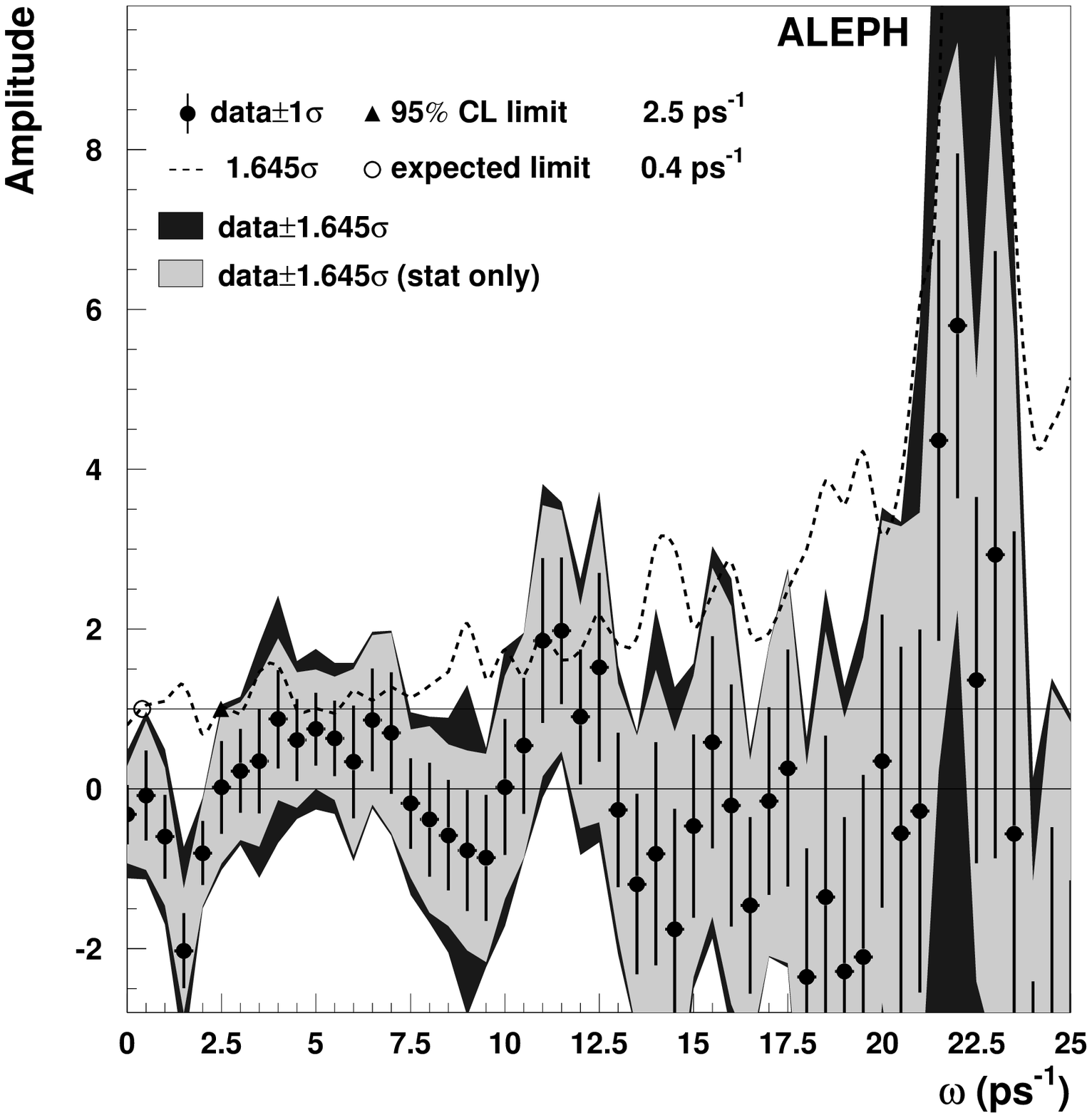,height=7.0cm}
  \caption{(left) Mass spectra for the $\PBs$ candidates in the fully exclusive
	analysis in data (dots) and Monte Carlo (histograms): a) $\PsDm\pi^+
	(\pi^0,\gamma)$ channel, b) $\PsDm\mathrm{a}_1^+(\gamma)$,
	c) the sum of both.
	(right) Amplitude fit for the fully exclusive analysis.}
  \label{fig:bsmass}
  \end{center}
\end{figure}

\subsection{Flavour Tagging}
The three analyses use a similar algorithm for the determination of the
flavour at production. In the event hemisphere opposite to the one
containing the $\PBs$ candidate, a neural network is used to discriminate
tracks coming from the primary and the secondary vertex and to compute
jet charges and primary and secondary vertex charges, which are combined with
the charge of lepton or kaon candidates (if present) in a single estimator.

In addition, information from the hemisphere containing the $\PBs$ candidate is
also used, in particular by calculating charge estimators with the
fragmentation
tracks and looking at fragmentation kaon candidates. Finally, the
correlation between
the flavour and the angle of the $\PBs$ with respect to the initial electron
direction, due to the b forward-backward asymmetry, is exploited.

The average probability for signal events to tag incorrectly the initial
state flavour is found to be about 24\% for the three analyses.

The flavour state at decay is determined from the sign of the charge of the
$\PBs$ decay products. In the analyses using semileptonic decays, the
opposite correlation between the lepton charge and the $\PBs$ flavour
in $\mathrm{b}\rightarrow\mathrm{c}\rightarrow\ell$ cascade decays with
respect to the signal is taken into account.

\begin{figure}
  \begin{center}
    \epsfig{file=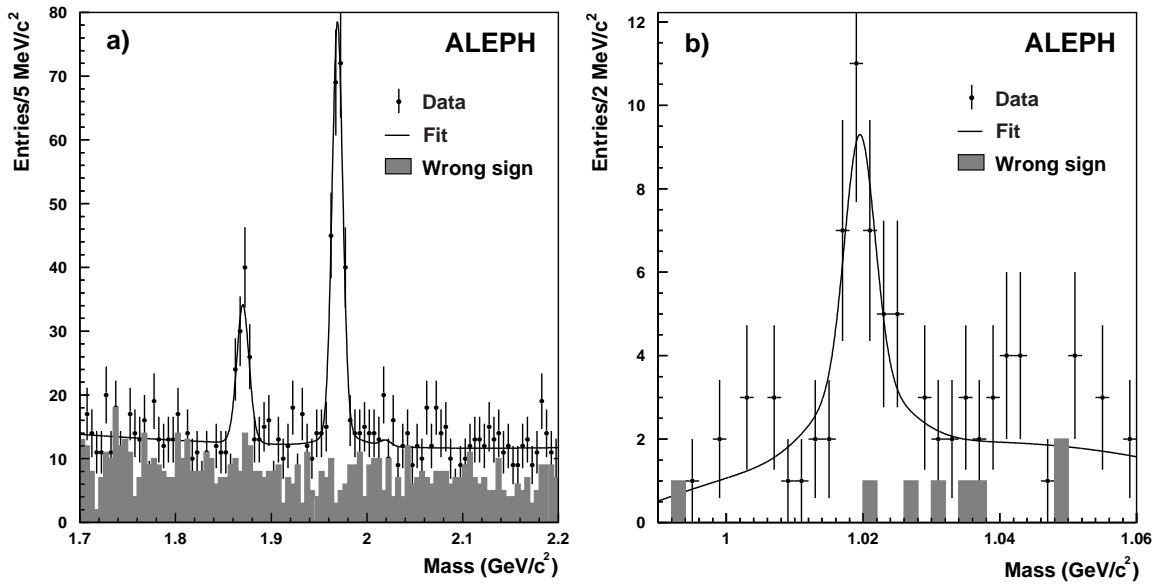,height=8.5cm}
    \caption{Mass distribution of candidates with $\PsDm\ell^+$ combinations
	for a) hadronic and b) semileptonic $\PsDm$ decays on data (dots)
	and Monte Carlo (histograms). The spectra for the wrong sign
	combinations are also shown.}
    \label{fig:dsm}
  \end{center}
\end{figure}

\subsection{Analysis of Fully Reconstructed Hadronic Decays}
This analysis is performed on a selected sample of $\PBs\rightarrow
\PsDstm\pi^+$, $\PBs\rightarrow\PsDstm\mathrm{a}_1^+$, and
$\PBs\rightarrow\PsDstm\rho^+$ decays, where
$\mathrm{D}^{\ast-}\rightarrow\PsDm\gamma$ and the $\PsDm$ meson
decays into $\Phi\pi^-$, $\PKstz\PKm$, or
$\PKzS\PKm$. The resolution of the reconstructed $\PBs$ mass is about 20
\mevcc, and the selection efficiency ranges from 5\% to 20\%, depending on the
decay channel.

The signal-to-background discrimination is largely improved by the
reconstruction of the photons and neutral pions emitted in the decays with
a $\mathrm{D}^{\ast-}$ or a $\rho^+$; otherwise, such decays would
be reconstructed with
a significantly lower mass and a much poorer mass resolution.

The selected candidates are 80, and their mass spectrum is shown in Figure
\ref{fig:bsmass}. The composition of the sample in terms of signal
$\PBs$ events, events with other b hadrons and lighter quark events, is
estimated on an event-by-event basis as a function of the helicity angle or
the invariant mass of the $\PsDm$ candidate with a standard discrimination
variable technique. The estimated signal purity is in average about 36\% but
can reach 80\% for some events.

The main interest of this analysis resides in the very precise measurement
of the decay proper time, due mainly to the fact that all the decay particles
are measured and therefore the $\PBs$ momentum is accurately known, at the
percent level. The error on the measured decay length is around 180 $\mu$m,
and the error on the proper time is about 0.08 ps. This results
in a relatively slower degradation of the amplitude error at increasing test
frequencies, and allows this analysis to give a significant contribution to
the combined amplitude spectrum at high frequencies, despite its very low
sensitivity (Figure \ref{fig:bsmass}).

\subsection{Analysis of $\PsDm\ell^+$ Pairs}
The second analysis discussed here relies on
the reconstruction of $\PBs\rightarrow\PsDstm\ell^+\nu_\ell$
decays, with the $\PsDm$ meson decaying in one of the following hadronic
and semileptonic decay modes:
\begin{eqnarray}
  \begin{array}{ll}
    \nonumber
    \PsDm\to\phi\,\pi^- \,,&
    \PsDm\to\PKstz\,\PKm \,,\\
    \PsDm\to\PKzS\,\PKm \,, &
    \PsDm\to\phi\,\rho^-\,, \\
    \PsDm\to\PKstz\,\PKst^-\,,&
    \PsDm\to\phi\,\pi^+\,\pi^-\,\pi^-\,,\\
    \PsDm\to\phi\,\Pem\,\bar{\nu}_{\mathrm{e}} \,,&
    \PsDm\to\phi\,\mu^-\,\bar{\nu}_{\mu}~.
  \end{array}
\end{eqnarray}

Compared with a similar analysis previously
published~\cite{oldbs}, it takes advantage from an improved $dE/dx$
estimation and tracking performance, a general reoptimization of the
selection cuts and a better background discrimination.

As the neutrino from the $\PBs$ decay (and the one from the $\PsDm$ decay,
if any) is undetected, its energy is estimated from the missing energy in the
hemisphere~\cite{eneu}. This results in an  average relative uncertainty on the
$\PBs$ momentum of about 11\%. The decay length uncertainty
is about 240 $\mu$m and it is
worse than for the fully exclusive analysis due to the smaller angle
among the decay particles in the laboratory system.

The candidates selected are 333; a major source of background to the event
selection comes from
$\mathrm{b}\rightarrow\PsDm\PD\mathrm{X}\ (\mathrm{D}\rightarrow
\ell^+)$ decays, which can be discriminated from the signal by the differences
in shape of the distributions of the lepton momentum,
the $\PsDm\ell^+$ invariant mass, the estimated $\PBs$ momentum and the number
of charged particles forming a good vertex with the lepton. The fraction
of combinatorial background is estimated from a fit to the
$\PsD$ ($\phi$) mass spectra for the hadronic (semileptonic) decays, shown
in Figure \ref{fig:dsm}. The signal purity depends strongly on the decay
mode, and is 47\% in average.

\begin{figure}
  \begin{center}
    \epsfig{file=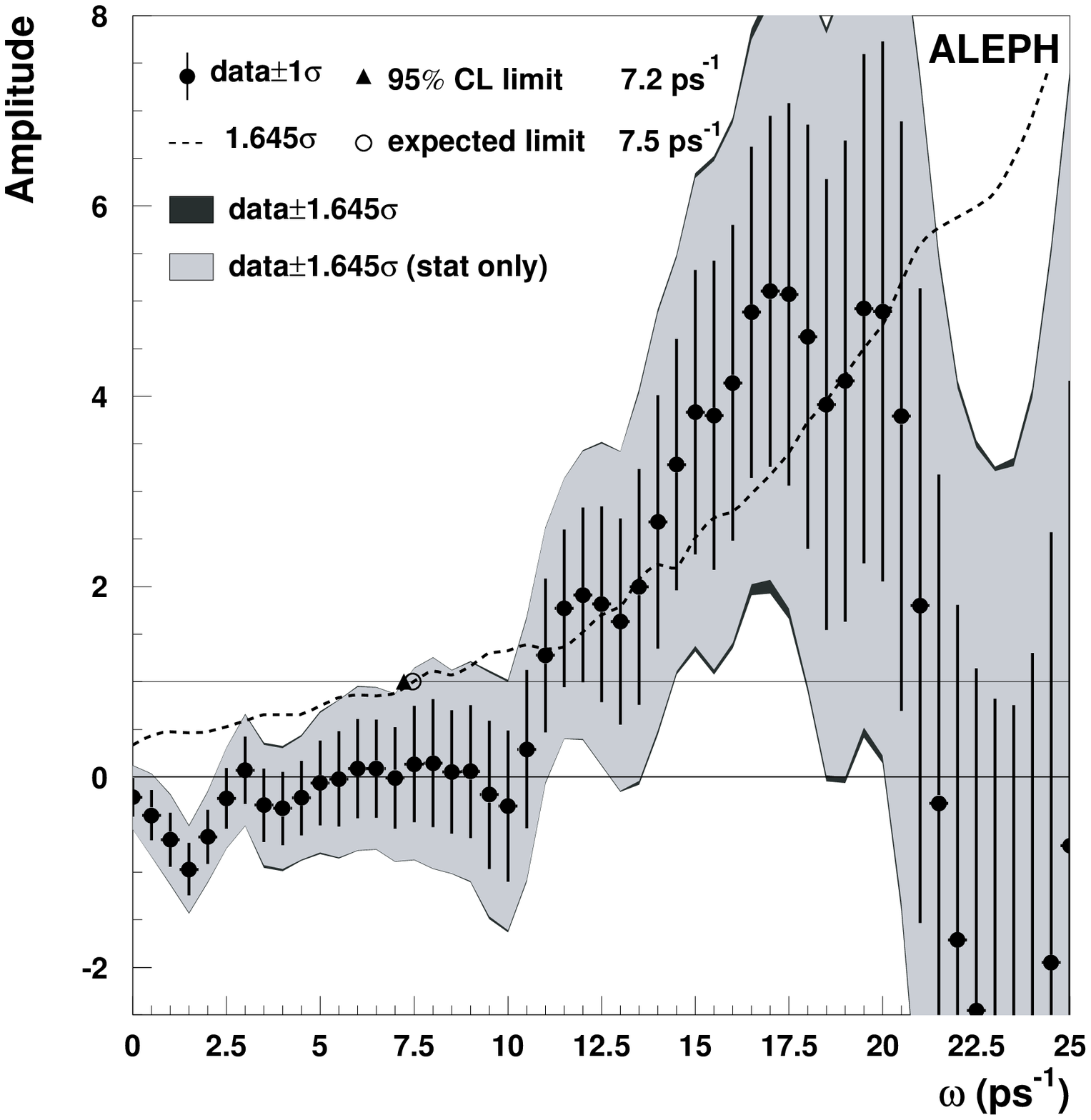,height=7.9cm}
    \epsfig{file=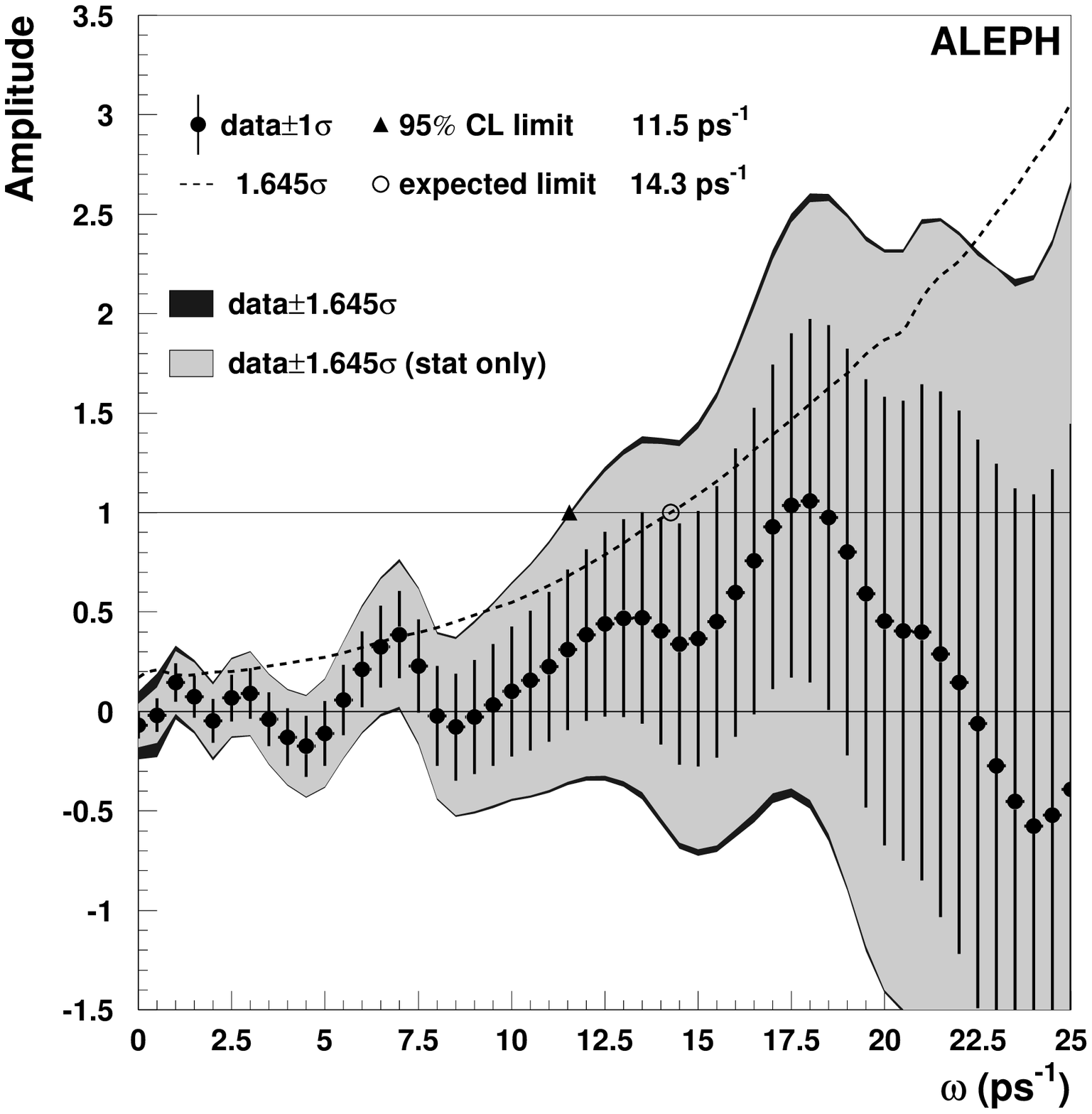,height=7.9cm}
    \caption{Amplitude fit as a function of the frequency for the
	$\PsDm\ell^+$ analysis (left) and for the inclusive semileptonic
	analysis (right).}
    \label{fig:ampl_aleph}
  \end{center}
\end{figure}

The result of the fit to the amplitude of the oscillation term is shown
in Figure \ref{fig:ampl_aleph}.

\subsection{Inclusive Semileptonic Analysis}
This analysis is based on
a sample of inclusively selected $\PBs$ semileptonic decays, and supersedes
a similar analysis~\cite{oldincl}. Event hemispheres
with a lepton having a large transverse momentum relative to the nearest jet
are selected, and the charmed particle from the b-hadron decay is reconstructed
by means of a topological algorithm. The inclusion
in the secondary vertex fit of a ``B track'' together with the lepton and the
charmed particle, and the reconstruction of photons from the decay of the
charmed particle, allow to improve the $\PBs$ decay length measurement
precision up to 22\%.

In average, the decay length resolution is about 370 $\mu$m; however,
the statistical significance of the data sample is greatly enhanced by
parameterizing the decay length resolution as a function of several
quantities, mostly related to the topology of the decay and the quality
of the vertex reconstruction. 
The $\PBs$ momentum is estimated using the lepton energy, the neutrino
energy, and the energy of a jet clustered around the charged
particles coming from
the D vertex. The momentum resolution is parameterized event-by-event, and
is about 12\% in average.

The fraction of $\mathrm{b}\overline{\mathrm{b}}$ events in the selected
sample is enlarged by the use of a b-tagging
algorithm, and a neural network is used to discriminate direct $\mathrm{b}
\rightarrow\ell$ decays with respect to cascade decays ($\mathrm{b}
\rightarrow\mathrm{c}\rightarrow\ell$), exploiting the differences
between them in terms of lepton kinematics and vertex topology. Finally,
the abundance of $\PBs$ decays compared to other b hadron species is estimated
as a function of the total charge and the charged multiplicity of the tertiary
vertex, and of the presence of kaons coming from the fragmentation and the
decay of the charmed particle. After the selection, 74026 candidates remain, of
which 87\% contain $\mathrm{b}\rightarrow\ell$ decays, and 10\% are
$\PBs$ decays.

The results of the amplitude fit performed on this sample are visible in
Figure \ref{fig:ampl_aleph}.
It is worth to stress that, compared with the previous version of this
analysis, the event statistics is more that two times larger, and the
description of the proper time resolution and the signal purity is much
more detailed, resulting in an increase in the sensitivity from 9.6 ps$^{-1}$
to 14.3 ps$^{-1}$.

\subsection{Combined Result}
\begin{figure}[t]
  \begin{center}
    \epsfig{file=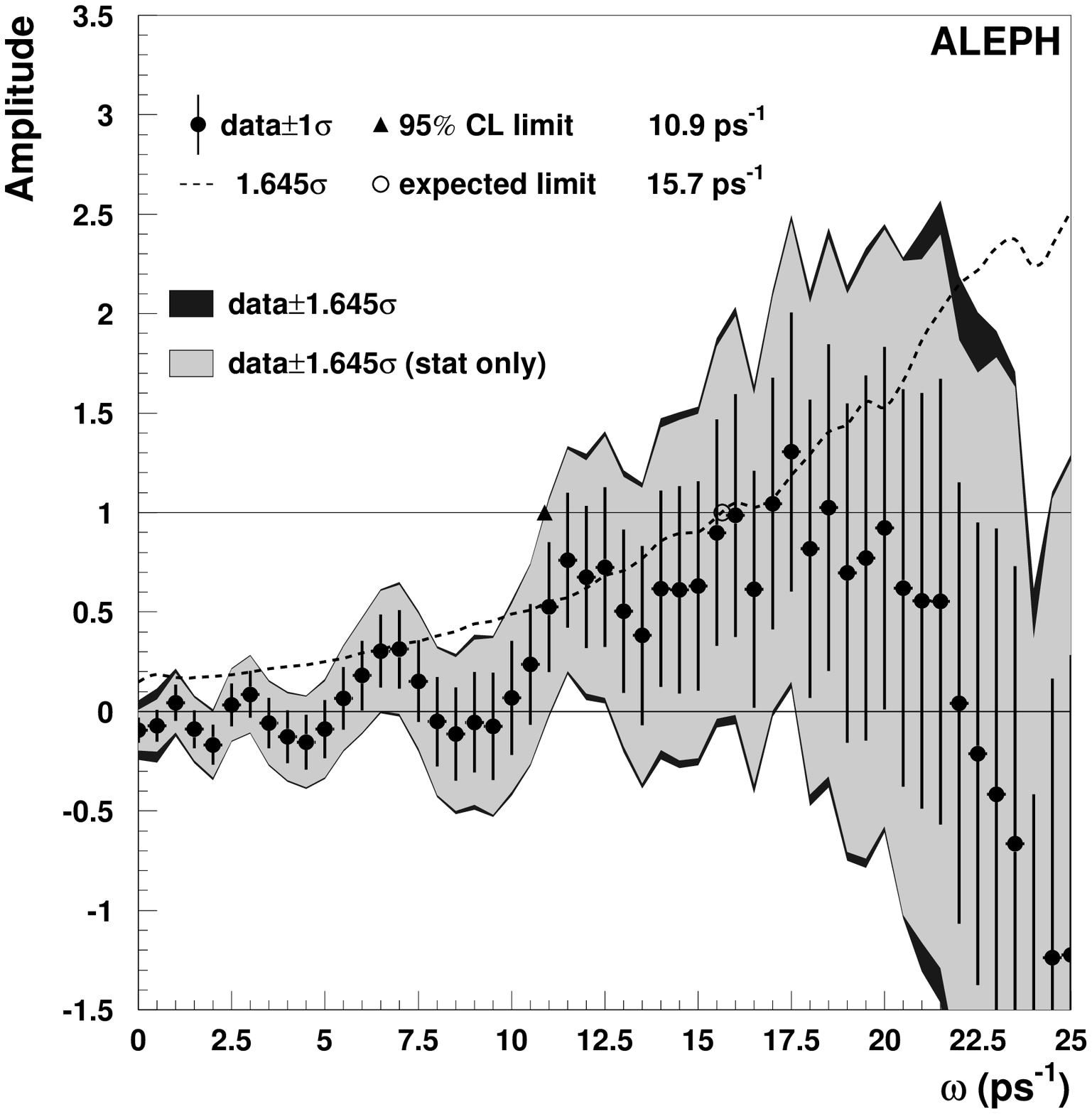,height=7.9cm}
    \epsfig{file=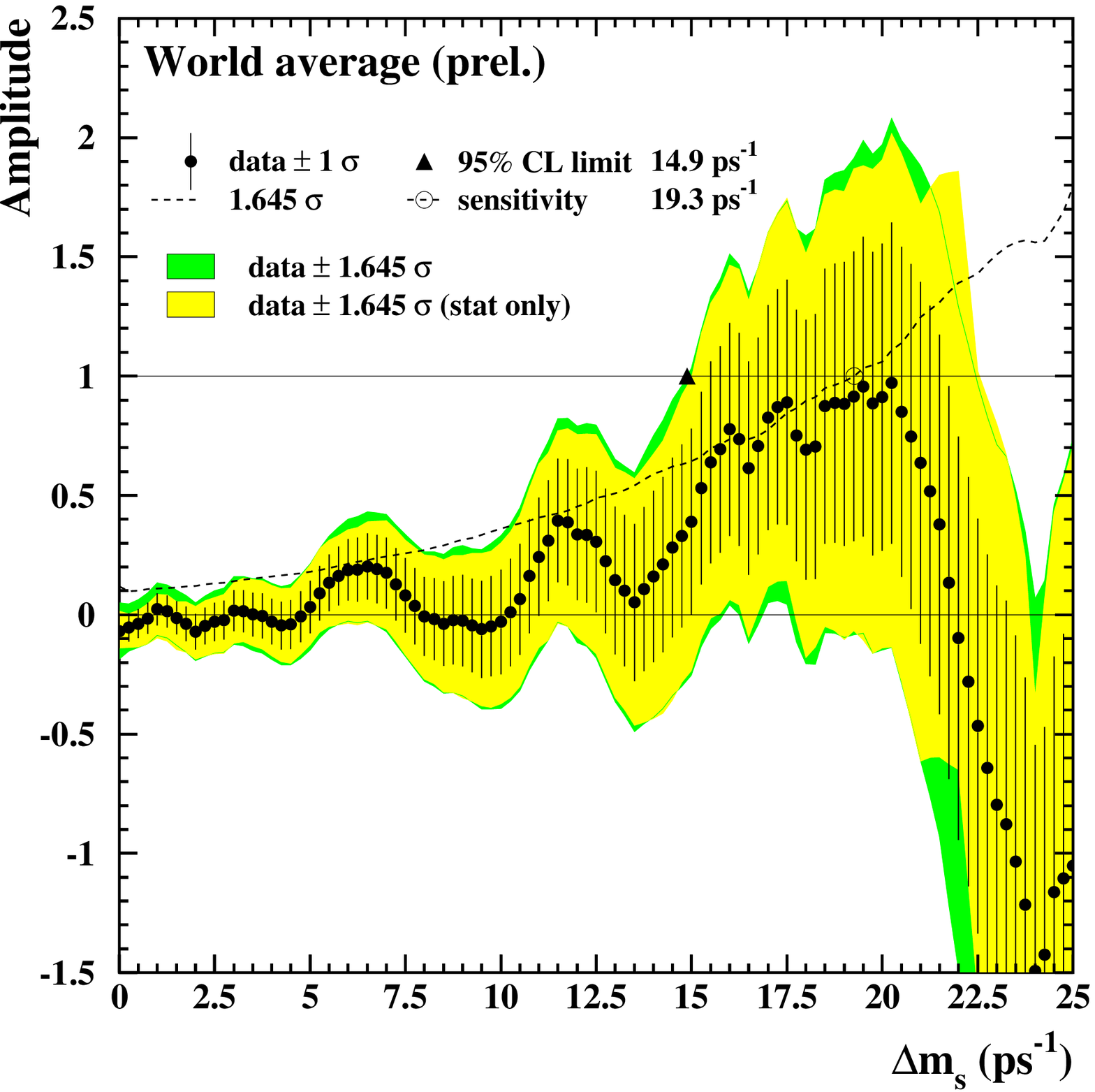,height=7.9cm}
    \caption{(left) ALEPH combined amplitude fit. (right) Amplitude fit
	for the world combination.}
    \label{fig:worldfit}
  \end{center}
\end{figure}

The amplitude spectra for all three analyses have been combined and the result
is shown in Figure \ref{fig:worldfit}. It can be seen that a 95\% C.L.
lower limit of 10.9 ps$^{-1}$ can be put on the $\PBs$ oscillation
frequency, compared to a sensitivity of 15.7 ps$^{-1}$. This is,
presently, the highest sensitivity reached by a single experiment.

\section{Present Status of the $\dms$ Measurements}
The new ALEPH results have been combined with all the available results
from other experiments (some of which still preliminary)~\cite{comb},
and the amplitude
spectrum for the combination is shown in Figure~\ref{fig:worldfit}.
The 95\% C.L. lower limit on $\dms$ is 14.9 ps$^{-1}$, and the
sensitivity reaches 19.3 ps$^{-1}$. These numbers are within the range allowed
by a global fit on the unitarity triangle parameters~\cite{triangle}, which
gives $\dms=17.8^{+3.2}_{-2.8}$ ps$^{-1}$.

It is clear that the final sensitivity to be reached by all the analyses
from the LEP collaborations and SLD combined together, cannot be far from
its present value. On the other hand, the Tevatron Run II data is expected
to allow CDF and D0 to reach, by the 2002 summer, a sensitivity of several
tens of ps$^{-1}$, well beyond the value given by the
indirect measurement quoted above.

\end{document}